\documentclass[aps,prd,showpacs,notitlepage,nofootinbib,preprintnumbers,amsmath,amssymb]{revtex4-1}

\usepackage{graphics,graphicx}
\usepackage{dcolumn}
\usepackage{bm}
\usepackage{epsfig}
\usepackage[usenames]{color}
\usepackage{hyperref} 
\usepackage{mathbbol}
\usepackage{mathtools}
\usepackage{epstopdf}
\usepackage{simplewick}
\usepackage[utf8]{inputenc} 	
\usepackage{slashed}
\usepackage{stackrel}
\usepackage[normalem]{ulem}

\newcommand{\ul}[1]{\underline{#1}}

\newcommand{\ord}[1]{\mathcal{O}\left( #1 \right)}

\newcommand{\half}{\frac{1}{2}}

\begin{document}
\title{A complete set of in-medium  splitting functions  to any order in opacity }
\author{Matthew D. Sievert}
  \email[Email: ]{matthew.sievert@rutgers.edu}
  \affiliation{Department of Physics and Astronomy, Rutgers University, Piscataway, NJ 08854, USA}
\author{Ivan Vitev}
  \email[Email: ]{ivitev@lanl.gov}
  \affiliation{Theoretical Division, Los Alamos National Laboratory, Los Alamos, NM 87545, USA}
\author{Boram  Yoon}
  \email[Email: ]{boram@lanl.gov}
  \affiliation{ Computer, Computational, and Statistical Sciences Division, Los Alamos National Laboratory, Los Alamos, NM 87545, USA}
\date{\today}
%
%
\begin{abstract}

In this Letter we report the first calculation of all ${\cal O}(\alpha_s)$  medium-induced  branching  processes  to  any order in opacity.  Our splitting functions results are  presented as iterative  solutions to matrix equations with initial conditions set by the leading order  branchings in the vacuum. The flavor and  quark mass  dependence of the in-medium $q \rightarrow qg$, $g\rightarrow gg$, $q \rightarrow g	q$, $g \rightarrow q\bar{q}$ processes is fully  captured by the light-front wavefunction formalism and the color representation of the parent and daughter partons.  We include the explicit solutions to second order in opacity as supplementary material and present numerical results in a realistic strongly-interacting  medium produced in high center-of-mass  energy heavy ion  collisions at the Large Hadron Collider. Our numerical simulations show that the second order in opacity corrections can change  the  energy dependence of the in-medium shower intensity.  We further find corrections to the longitudinal and angular  distributions of the in-medium splitting kernels  that  may have important implications for jet substructure phenomenology.   
\end{abstract}
%
%

%
\maketitle
%

%
\section{Introduction}
\label{intro}
%

In Quantum Chromodynamics (QCD), the probability of one parton to split into a two-parton system is governed by the well-known Altarelli-Parisi splitting functions in vacuum~\cite{Altarelli:1977zs}. These probabilities are the key ingredients in all modern high-precision calculations in the theory of strong interactions and in Monte-Carlo event generators for high energy physics. For jet physics, quark and gluon splitting kernels play an essential role in  understanding the radius dependence of inclusive cross sections and of jet substructure. These can be directly measured using novel observables, such as the 
groomed soft-dropped momentum sharing distribution of the two leading subjets inside a reconstructed jet~\cite{Larkoski:2017bvj}.  For jets propagating through hot or cold QCD matter, the vacuum splitting functions receive medium-induced contributions~\cite{Wang:2009qb,Ovanesyan:2011xy,Ovanesyan:2011kn,Blaizot:2012fh,Fickinger:2013xwa,Apolinario:2014csa,Ovanesyan:2015dop,Kang:2016ofv}  where non-local destructive interference, known as the non-Abelian Landau-Pomeranchuk-Migdal (LPM) effect~\cite{Wang:1994fx,Zakharov:1996fv,Baier:1996kr,Gyulassy:2000fs,Wang:2001ifa,Arnold:2002ja}, plays an important role.  These modifications can then be calculated order by order in powers of the opacity $\chi$, the average number of scatterings in the medium.  The basis of the opacity expansion and the current state of the field are reviewed in Ref.~\cite{Sievert:2018imd} and the references therein.

So far, a realistic numerical evaluation of higher order-in-opacity corrections has been missing from the literature. Early works investigated the convergence of the opacity series to second and third order, utilizing the soft gluon emission limit available at the time~\cite{Gyulassy:2000fs,Gyulassy:2000er}.  These studies used models of the QCD medium density carefully chosen to smooth out the non-Abelian LPM interference pattern, making certain integrals analytically calculable.  It was found that for opacities as large as five  the series converges quickly and the corrections to parton energy loss can be of the order of a few percent, especially at  typical jet energies available at the LHC. The energy spectrum of radiated gluons receives larger corrections, of order 10 - 30\%,  and the fully differential angular distribution  of the radiation can be affected even more. Similar conclusions were reached in  Ref.~\cite{Wicks:2008ta},  where corrections up to ninth order in opacity were  evaluated, albeit with much reduced numerical accuracy. In the high-twist approach second order corrections were calculated in Ref.~\cite{Guo:2006kz}. The possibility of a hard scattering was added to the  infinite medium soft scattering soft scattering approximation~\cite{Mehtar-Tani:2019tvy}.  Importantly, it was recently shown in the work of Feal and Vasquez~\cite{Feal:2018sml,Feal:2018jbm} that analytic approximations to in-medium splitting processes are not a good substitute for their full numerical evaluation.  

In Ref.~\cite{Sievert:2018imd} we constructed a new framework to compute these splitting functions to higher orders in the opacity expansion, which was previously limited to either the soft-gluon approximation $x \ll 1$ or to first order in opacity.  With the derivation of the full in-medium splitting functions and theoretical formalisms that bridge the gap between high energy and heavy ion physics, the question of corrections from higher orders in opacity must now be revisited.  We are also motivated by the more precise experimental  measurements of jets and jet substructure that have become available or are expected  in the near future, for recent examples see~\cite{Adamczyk:2017yhe,Zharko:2017vsl,Sirunyan:2018gct,Sirunyan:2018qec,Aaboud:2019oac,Sirunyan:2018ncy,Aaboud:2018twu,Aaboud:2018hpb}. In-medium splitting kernels enter {\em directly} into the fixed order and resummed calculation of such observables~\cite{Chien:2015hda,Chien:2016led,Kang:2017frl,Li:2017wwc,Li:2018xuv}.  In this Letter, we present new results which extend the formalism constructed in Ref.~\cite{Sievert:2018imd} to allow the calculation of all partonic splitting channels beyond the soft-gluon approximation to any order in opacity, and we evaluate them numerically at second order in opacity for the first time. Unlike the early works which used medium models chosen for theoretical convenience, here we employ realistic state-of-the-art hydrodynamic models of the expanding quark-gluon plasma to enable precision jet phenomenology.
%

For the purposes of this Letter we will limit ourselves to the final-state interactions of a jet produced in the medium, and we will work in the limit where the energy of the jet is much larger than any medium scales such that collisional energy loss is suppressed compared to radiative branching.  
Throughout this Letter we will use light-front coordinates $v^\pm \equiv \frac{1}{\sqrt{2}} (v^0 \pm v^3)$ and denote transverse vectors by $\ul{v} = (v_\bot^1 , v_\bot^2)$ with magnitudes $v_T \equiv | \ul{v} |$.  We work in a general frame in which the jet moves along the $+z$ axis with longitudinal momentum $p^+$; all the results are manifestly boost invariant, and the medium rest frame is just the special case for which $p^+ = \sqrt{2} E$ with $E$ the jet energy.  The rest of our Letter is organized as follows: in Section~\ref{analytic} we outline the analytic derivation of all medium-induced splitting functions. Their numerical evaluation is discussed in  Section~\ref{numeric}, and explicit new results to second order in opacity are given.  We give our conclusions in Section~\ref{conclusions}.

%
\section{Universal Features of Medium-Induced Parton Branching}
\label{analytic}
%

For the four different branching channels in QCD to lowest order, we will write $a \rightarrow b c$ with partons $a, b, c$ being either (anti)quarks or gluons.  We take the parent parton $a$ to have large longitudinal momentum $p^+$ along the jet  axis, with the daughter partons $b$ and $c$ carrying fractional momenta $(1-x) p^+$ and $x p^+$, respectively.  We also denote by $\ul{k}$ the relative transverse momentum given to parton $c$ by the branching.  As was shown for the 
$q \rightarrow  qg$ branching in Ref.~\cite{Sievert:2018imd}, the splitting functions can be expressed in terms of the following quantities: the light-front wave function (LFWF) of the branching $\psi$ \cite{Lepage:1980fj, Brodsky:1997de}, the corresponding energy denominator $\Delta E^-$, the cross-section $\frac{d\sigma^{el}}{d^2 q}$ to interact with a scattering center in the medium, and the scattering length $\lambda$ of a parton.  These features are universal to all partonic branching channels $a \rightarrow b c$ and to any QCD medium; the LFWF themselves will differ for different channels, but the way in which they are dressed by the medium is the same.

The LFWF (squared and averaged over quantum numbers) and energy denominator for any channel can be written in the form
\begin{subequations}
\begin{align} \label{e:LFWFsq}
\left\langle \psi(x , \ul{\kappa}) \, \psi^* (x , \ul{\kappa'}) \right\rangle &=
\frac{8\pi\alpha_s \: f(x)}{ [\kappa_T^2 + \nu^2 m^2] \, [\kappa_T^{\prime \, 2} + \nu^2 m^2]}
\bigg[ g(x) \:(\ul{\kappa} \cdot \ul{\kappa}') + \nu^4 m^2 \bigg] \, ,
\\
\Delta E^- (x , \ul{\kappa}) &= - \frac{ \kappa_T^2 + \nu^2 m^2}{2 x (1-x) p^+} \, , 
\end{align}
\end{subequations}
with functions $f(x), g(x)$ describing the distribution of longitudinal momentum in the branching and a coefficient $\nu$ accompanying the mass dependence.  These functions and coefficients will be fully listed in Table~\ref{tablecoeffs} below.  The LFWF describe the branching in vacuum, that is, at $0^{th}$ order in opacity $\ord{\chi^0}$:
\begin{align} \label{e:0thOrd}
\left. p^+ \frac{dN}{d^2 k \, dx \, d^2 p \, dp^+} \right|_{\ord{\chi^0}} &= 
\frac{\mathcal{C}_0}{2(2\pi)^3 \, x \, (1-x)} \:
\langle \psi(x , \ul{k - x p}) \, \psi^* (x , \ul{k - xp}) \rangle \, \times \,
\left( p^+ \frac{dN_0}{d^2 p \, dp^+} \right) ,
\end{align}
where $\mathcal{C}_0$ is the color factor for the vacuum splitting and $p^+ \frac{dN_0}{d^2p \, dp^+}$ is the distribution of the parent parton with momentum $(p^+ , \ul{p})$ created by the hard scattering in the medium.  As in Ref.~\cite{Sievert:2018imd}, each additional order in opacity corresponds to an additional correlated rescattering in the medium and generates a corresponding integral over both the position $z^+$ of and momentum exchange $\ul{q}$ with the scattering center:
%
$\int\frac{dz^+}{\lambda_R^+} \int\frac{d^2 q}{\sigma_{el}} \frac{d\sigma^{el}}{d^2 q}$,
%
where $\lambda_R^+$ is the mean free path of a parton in color representation $R$ ($\lambda_q^+$ for a quark or $\lambda_G^+$ for a gluon) and $\sigma_{el} = \int d^2 q \, \frac{d\sigma^{el}}{d^2 q}$ is the elastic scattering cross section on a scattering center.  Local color neutrality of the scattering centers implies that the leading-order scattering process involves two-gluon exchange at the level of the amplitude squared, corresponding to 9 ``direct" scattering diagrams and 8 ``virtual" scattering diagrams, labeled $D_{1-9}$ and $V_{1-8}$ respectively in the notation of Ref.~\cite{Sievert:2018imd}.  The color factors for each of these diagrams are specified by coefficients $d_{1-6}$ and $v_{1 , 3 , 5 , 7}$ that differ for each channel, but they are preserved and iterated at successive orders in opacity.
\footnote{The direct diagrams $D_7, D_8, D_9$ are the complex conjugates of diagrams $D_4, D_5, D_6$ and correspondingly enter with the same color factors $d_4, d_5, d_6$, respectively.  Likewise the virtual diagrams $V_2, V_4, V_6, V_8$ are complex conjugates of diagrams $V_1, V_3, V_5, V_7$ and enter with the same color factors $v_1, v_3, v_5, v_7$, respectively.}  

Each partonic splitting channel is then fully specified by the quantities summarized in Table~\ref{tablecoeffs}. 
\begin{table} 
\begin{align*}
\begin{array} {c | c c c c c c | c c c c | c c | c c c}
 & d_1 & d_2 & d_3 & d_4 & d_5 & d_6 & v_1 & v_3 & v_5 & v_7 & \lambda_R^+ 
& \mathcal{C}_0 & \nu & f(x) & g(x) 
\\ \hline
%
q \rightarrow qg
&  1 & 1 & \frac{N_c}{C_F} & \frac{-1}{2 N_c C_F} & \frac{N_c}{2 C_F} & \frac{- N_c}{2 C_F} &
-\half & -\half & \frac{-N_c}{2 C_F} & \frac{N_c}{2 C_F} & \lambda_q^+ & C_F & x & 1-x & 1 + (1-x)^2 
\\
%
q \rightarrow gq 
&  1 & 1 & \frac{N_c}{C_F} & \frac{-1}{2 N_c C_F} & \frac{N_c}{2 C_F} & \frac{- N_c}{2 C_F} &
-\half & -\half & \frac{-N_c}{2 C_F} & \frac{N_c}{2 C_F} & \lambda_q^+ & C_F & 1-x & x & 1 + x^2
\\
%
g \rightarrow q \bar{q} 
&  1 & \frac{C_F}{N_c} & \frac{C_F}{N_c} & \half & \half & \frac{1}{2 N_c^2} &
-\half & -\frac{C_F}{2 N_c} & \frac{-C_F}{2 N_c} & \frac{-1}{2 N_c^2} & \lambda_G^+ & \half &
1 & x (1-x) & x^2 + (1-x)^2
\\
%
g \rightarrow g g
&  1 & 1 & 1 & \half & \half & -\half & -\half & -\half & -\half & \half & \lambda_G^+ & N_c & 0 & 1 + x^4 + (1-x)^4 & 1
\end{array}
\end{align*}
\caption{Complete list of functions and coefficients for all four parton branching processes in QCD matter.}
\label{tablecoeffs}
\end{table}
The choice of representation $R$ used in the mean free path is arbitrary, but this choice rescales the scattering color factors $d_{1-6}$ and $v_{1 , 3 , 5 , 7}$; here we adopt the convention of using the representation of the parent parton for each channel.  Note that the $q \rightarrow gq$ channel is obtained from the $q \rightarrow qg$ channel under the mapping $x \rightarrow (1-x) \, , \, \ul{k} \rightarrow - \ul{k}$.  This transformation leaves the splitting function invariant as long as the cross section $\frac{d\sigma^{el}}{d^2 q}$ is even under $\ul{q} \rightarrow - \ul{q}$. For finite kinematics, we can expect some  deviations from this symmetry.    

As derived in Ref.~\cite{Sievert:2018imd}, the contribution from $N^{th}$ order in opacity to the distribution of partons within a jet can be expressed as
\begin{align} \label{e:NthOrd}
\left. p^+ \frac{dN}{d^2 k \, dx \, d^2 p \, dp^+} \right|_{\ord{\chi^N}} &= 
\frac{\mathcal{C}_0}{2(2\pi)^3 \, x \, (1-x)} \:
f^{(N)}_{F/F} (\ul{k} , \ul{k} , \ul{p} ; \infty^+ , \infty^+) .
\end{align}
The function  $f^{(N)}_{F/F} $ is computed using the generalized recursion relations  (referred to as the reaction operator)
\begin{subequations} \label{e:reacts}
\begin{align} \label{e:react1}
f_{F / F}^{(N)} & (\ul{k} , \ul{k}' , \ul{p} \, ; \, x^+ , y^+) = 
\int\limits_{0}^{\min[ x^+ , y^+ , L^+]} \hspace{-0.6cm} \frac{dz^+}{\lambda_R^+} \:
\int \frac{d^2q}{\sigma_{el}} \frac{d\sigma^{el}}{d^2 q}
\: \times \Bigg\{
\notag \\ & 
d_2 \,
e^{i [ \Delta E^- (\ul{k} - x \ul{p} + x \ul{q}) - \Delta E^- (\ul{k} - x \ul{p}) ] z^+} \:
e^{i [ \Delta E^- (\ul{k}' - x \ul{p}) - \Delta E^- (\ul{k}' - x \ul{p} + x \ul{q}) ] z^+}
f_{F / F}^{(N-1)} (\ul{k} , \ul{k}' , \ul{p} - \ul{q} \, ; \, z^+ , z^+)
\notag \\ +&
d_3 \, 
e^{i [ \Delta E^- (\ul{k} - x \ul{p} - (1-x) \ul{q}) - \Delta E^- (\ul{k} - x \ul{p}) ] z^+} \:
e^{i [ \Delta E^- (\ul{k}' - x \ul{p}) - \Delta E^- (\ul{k}' - x \ul{p} - (1-x) \ul{q}) ] z^+}
f_{F / F}^{(N-1)} (\ul{k} - \ul{q} , \ul{k}' - \ul{q} , \ul{p} - \ul{q} \, ; \, z^+ , z^+)
\notag \\ +&
d_6 \,
e^{i [ \Delta E^- (\ul{k} - x \ul{p} + x \ul{q}) - \Delta E^- (\ul{k} - x \ul{p}) ] z^+} \:
e^{i [ \Delta E^- (\ul{k}' - x \ul{p}) - \Delta E^- (\ul{k}' - x \ul{p} - (1-x) \ul{q}) ] z^+}
f_{F / F}^{(N-1)} (\ul{k} , \ul{k}' - \ul{q} , \ul{p} - \ul{q} \, ; \, z^+ , z^+)
\notag \\ +&
d_6 \,
e^{i [ \Delta E^- (\ul{k} - x \ul{p} - (1-x) \ul{q}) - \Delta E^- (\ul{k} - x \ul{p}) ] z^+} \:
e^{i [ \Delta E^- (\ul{k}' - x \ul{p}) - \Delta E^- (\ul{k}' - x \ul{p} + x \ul{q}) ] z^+}
f_{F / F}^{(N-1)} (\ul{k} - \ul{q} , \ul{k}' , \ul{p} - \ul{q} \, ; \, z^+ , z^+)
   \notag \\ +& 
(v_3 + v_5) \, f_{F / F}^{(N-1)} (\ul{k} , \ul{k}' , \ul{p} \, ; \, z^+ , y^+)
+ (v_3 + v_5)  \, f_{F / F}^{(N-1)} (\ul{k} , \ul{k}' , \ul{p} \, ; \, x^+ , z^+)
\notag \\ +&
v_7 \,
e^{i [ \Delta E^- (\ul{k} - x \ul{p} - \ul{q}) - \Delta E^- (\ul{k} - x \ul{p}) ] z^+} 
f_{F / F}^{(N-1)} (\ul{k} - \ul{q} , \ul{k}' , \ul{p} \, ; \, z^+ , y^+)
 \notag \\ +&
 v_7
e^{i [ \Delta E^- (\ul{k}' - x \ul{p}) - \Delta E^- (\ul{k}' - x \ul{p} - \ul{q}) ] z^+} 
f_{F / F}^{(N-1)} (\ul{k} , \ul{k}' - \ul{q} , \ul{p} \, ; \, x^+ , z^+)
\notag \\ +&
d_4 \, \psi(\ul{k} - x \ul{p}) 
\left[ e^{-i \Delta E^- (\ul{k} - x \ul{p}) x^+} - e^{-i \Delta E^- (\ul{k} - x \ul{p}) z^+} \right]
\, e^{i [ \Delta E^- (\ul{k}' - x \ul{p}) - \Delta E^- (\ul{k}' - x \ul{p} + x \ul{q}) ] z^+} 
f_{I / F}^{(N-1)} (\ul{k}'  , \ul{p} - \ul{q} \, ; \, z^+ , z^+)
\notag \\ +&
d_5 \, \psi(\ul{k} - x \ul{p}) 
\left[ e^{-i \Delta E^- (\ul{k} - x \ul{p}) x^+} - e^{-i \Delta E^- (\ul{k} - x \ul{p}) z^+} \right]
\, e^{i [ \Delta E^- (\ul{k}' - x \ul{p}) - \Delta E^- (\ul{k}' - x \ul{p} - (1-x) \ul{q}) ] z^+} 
f_{I / F}^{(N-1)} (\ul{k}' - \ul{q} , \ul{p} - \ul{q} \, ; \, z^+ , z^+)
\notag \\ +&
v_1 \, \psi(\ul{k} - x \ul{p}) 
\left[ e^{-i \Delta E^- (\ul{k} - x \ul{p}) x^+} - e^{-i \Delta E^- (\ul{k} - x \ul{p}) z^+} \right]
f_{I / F}^{(N-1)} (\ul{k}'  , \ul{p} \, ; \, z^+ , y^+)
\notag \\ +&
d_4 \, \psi^*(\ul{k}' - x \ul{p}) 
\left[ e^{i \Delta E^- (\ul{k}' - x \ul{p}) y^+} - e^{i \Delta E^- (\ul{k}' - x \ul{p}) z^+} \right]
\, e^{i [ \Delta E^- (\ul{k} - x \ul{p} + x \ul{q}) - \Delta E^- (\ul{k} - x \ul{p}) ] z^+} 
f_{F / I}^{(N-1)} (\ul{k}  , \ul{p} - \ul{q} \, ; \, z^+ , z^+)
\notag \\ +&
d_5 \, \psi^*(\ul{k}' - x \ul{p}) 
\left[ e^{i \Delta E^- (\ul{k}' - x \ul{p}) y^+} - e^{i \Delta E^- (\ul{k}' - x \ul{p}) z^+} \right]
\, e^{i [ \Delta E^- (\ul{k} - x \ul{p} - (1-x) \ul{q}) - \Delta E^- (\ul{k} - x \ul{p}) ] z^+} 
f_{F / I}^{(N-1)} (\ul{k} - \ul{q} , \ul{p} - \ul{q} \, ; \, z^+ , z^+)
\notag \\ +&
v_1 \,
\psi^*(\ul{k}' - x \ul{p})
\left[ e^{i \Delta E^- (\ul{k}' - x \ul{p}) y^+} - e^{i \Delta E^- (\ul{k}' - x \ul{p}) z^+} \right]
f_{F / I}^{(N-1)} (\ul{k} , \ul{p} \, ; \, x^+ , z^+)
\notag \\ +&
d_1 \, \psi(\ul{k} - x \ul{p})
\left[ e^{-i \Delta E^- (\ul{k} - x \ul{p}) x^+} - e^{-i \Delta E^- (\ul{k} - x \ul{p}) z^+} \right]
\left[ e^{i \Delta E^- (\ul{k}' - x \ul{p}) y^+} - e^{i \Delta E^- (\ul{k}' - x \ul{p}) z^+} \right]
\psi^*(\ul{k}' - x \ul{p})
\notag \\ & \times 
f_{I / I}^{(N-1)} (\ul{p} - \ul{q} \, ; \, z^+ , z^+) \Bigg\}\, ,
\end{align}
\begin{align} \label{e:react2}
f_{I / F}^{(N)} (\ul{k}' , \ul{p} \, ; \, x^+ , y^+) &= 
\int\limits_{0}^{\min[ x^+ , y^+ , L^+]} \hspace{-0.6cm} \frac{dz^+}{\lambda_R^+} \:
\int \frac{d^2q}{\sigma_{el}} \frac{d\sigma^{el}}{d^2 q}
\: \times \Bigg\{
\notag \\ +&
d_4 \, e^{i [ \Delta E^- (\ul{k}' - x \ul{p}) - \Delta E^- (\ul{k}' - x \ul{p} + x \ul{q}) ] z^+} 
f_{I / F}^{(N-1)} (\ul{k}'  , \ul{p} - \ul{q} \, ; \, z^+ , z^+)
\notag \\ +&
d_5 \, e^{i [ \Delta E^- (\ul{k}' - x \ul{p}) - \Delta E^- (\ul{k}' - x \ul{p} - (1-x) \ul{q}) ] z^+}
f_{I / F}^{(N-1)} (\ul{k}' - \ul{q} , \ul{p} - \ul{q} \, ; \, z^+ , z^+)
\notag \\ +&
v_1 \,
f_{I / F}^{(N-1)} (\ul{k}'  , \ul{p} \, ; \, z^+ , y^+)
+ (v_3 + v_5) \, f_{I / F}^{(N-1)} (\ul{k}' , \ul{p} \, ; \, x^+ , z^+)
\notag \\ +&
v_7 \, e^{i [ \Delta E^- (\ul{k}' - x \ul{p}) - \Delta E^- (\ul{k}' - x \ul{p} - \ul{q}) ] z^+} 
f_{I / F}^{(N-1)} (\ul{k}' - \ul{q} , \ul{p} \, ; \, x^+ , z^+)
\notag \\ +&
d_1 \, \psi^*(\ul{k}' - x \ul{p})
\left[ e^{i \Delta E^- (\ul{k}' - x \ul{p}) y^+} - e^{i \Delta E^- (\ul{k}' - x \ul{p}) z^+} \right]
f_{I / I}^{(N-1)} (\ul{p} - \ul{q} \, ; \, z^+ , z^+)
\notag \\ +&
v_1 \, \psi^*(\ul{k}' - x \ul{p})
\left[ e^{i \Delta E^- (\ul{k}' - x \ul{p}) y^+} - e^{i \Delta E^- (\ul{k}' - x \ul{p}) z^+} \right]
f_{I / I}^{(N-1)} (\ul{p} \, ; \, x^+ , z^+)
\Bigg\} \, ,
\end{align}
\begin{align} \label{e:react3}
f_{F / I}^{(N)} (\ul{k} , \ul{p} \, ; \, x^+ , y^+) &= 
\int\limits_{0}^{\min[ x^+ , y^+ , L^+]} \hspace{-0.6cm} \frac{dz^+}{\lambda_R^+} \:
\int \frac{d^2q}{\sigma_{el}} \frac{d\sigma^{el}}{d^2 q}
\: \times \Bigg\{
\notag \\ +&
d_4 \, e^{i [ \Delta E^- (\ul{k} - x \ul{p} + x \ul{q}) - \Delta E^- (\ul{k} - x \ul{p}) ] z^+} 
f_{F / I}^{(N-1)} (\ul{k}  , \ul{p} - \ul{q} \, ; \, z^+ , z^+)
\notag \\ +&
d_5 \, e^{i [ \Delta E^- (\ul{k} - x \ul{p} - (1-x) \ul{q}) - \Delta E^- (\ul{k} - x \ul{p}) ] z^+} 
f_{F / I}^{(N-1)} (\ul{k} - \ul{q} , \ul{p} - \ul{q} \, ; \, z^+ , z^+)
\notag \\ +&
v_1 \, f_{F / I}^{(N-1)} (\ul{k} , \ul{p} \, ; \, x^+ , z^+)
+. (v_3 + v_5) \, f_{F / I}^{(N-1)} (\ul{k} , \ul{p} \, ; \, z^+ , y^+)
\notag \\ +&
v_7 \, e^{i [ \Delta E^- (\ul{k} - x \ul{p} - \ul{q}) - \Delta E^- (\ul{k} - x \ul{p}) ] z^+} 
f_{F / I}^{(N-1)} (\ul{k} - \ul{q} , \ul{p} \, ; \, z^+ , y^+)
\notag \\ +&
d_1 \, \psi(\ul{k} - x \ul{p})
\left[ e^{-i \Delta E^- (\ul{k} - x \ul{p}) x^+} - e^{-i \Delta E^- (\ul{k} - x \ul{p}) z^+} \right]
f_{I / I}^{(N-1)} (\ul{p} - \ul{q} \, ; \, z^+ , z^+) 
\notag \\ +&
v_1 \, \psi(\ul{k} - x \ul{p}) 
\left[ e^{-i \Delta E^- (\ul{k} - x \ul{p}) x^+} - e^{-i \Delta E^- (\ul{k} - x \ul{p}) z^+} \right]
f_{I / I}^{(N-1)} (\ul{p} \, ; \, z^+ , y^+)
\Bigg\} \, ,
\end{align}
\begin{align} \label{e:react4}
f_{I / I}^{(N)} (\ul{p} \, ; \, \min[x^+ , y^+]) &= \hspace{-0.6cm}
\int\limits_{0}^{\min[ x^+ , y^+ , L^+]} \hspace{-0.6cm} \frac{dz^+}{\lambda_R^+} \:
\int \frac{d^2q}{\sigma_{el}} \frac{d\sigma^{el}}{d^2 q}
\: \Bigg[ d_1 \, f_{I / I}^{(N-1)} (\ul{p} - \ul{q} \, ; \, z^+) + 2 v_1 \,
f_{I / I}^{(N-1)} (\ul{p} \, ; \, z^+) \Bigg] \, .
\end{align}
\end{subequations}
The initial conditions for the functions $f^{(N)}_{F/F}$, $f^{(N)}_{I/F}$, $f^{(N)}_{F/I}$, $f^{(N)}_{I/I}$ are as follows
\begin{subequations} \label{e:initconds}
	\begin{align}
	f_{F / F}^{(0)} (\ul{k} , \ul{k}' , \ul{p} \, ; \, x^+ , y^+) &=
	\psi(\ul{k} - x \ul{p})
	\left[ e^{-i \Delta E^- (\ul{k} - x \ul{p}) x^+} - e^{-i \Delta E^- (\ul{k} - x \ul{p}) x_0^+} \right]
	\notag \\ & \hspace{1cm} \times
	\left[ e^{i \Delta E^- (\ul{k}' - x \ul{p}) y^+} - e^{i \Delta E^- (\ul{k}' - x \ul{p}) x_0^+} \right]
	\psi^*(\ul{k}' - x \ul{p})
	\left( p^+ \frac{dN_0}{d^2 p \, dp^+} \right) \, , 
	\\ 
	f_{I / F}^{(0)} (\ul{k}' , \ul{p} \, ; \, x^+ , y^+) &=
	\psi^*(\ul{k}' - x \ul{p})
	\left[ e^{i \Delta E^- (\ul{k}' - x \ul{p}) y^+} - e^{i \Delta E^- (\ul{k}' - x \ul{p}) x_0^+} \right]
	\left( p^+ \frac{dN_0}{d^2 p \, dp^+} \right)\, , 
	\\ 
	f_{F / I}^{(0)} (\ul{k} , \ul{p} \, ; \, x^+ , y^+) &= 
	\psi(\ul{k} - x \ul{p})
	\left[ e^{-i \Delta E^- (\ul{k} - x \ul{p}) x^+} - e^{-i  \Delta E^- (\ul{k} - x \ul{p}) x_0^+} \right]
	\left( p^+ \frac{dN_0}{d^2 p \, dp^+} \right) \, , 
	\\ 
	f_{I / I}^{(0)} (\ul{p} \, ; \, x^+ , y^+) &= \left( p^+ \frac{dN_0}{d^2 p \, dp^+} \right) \, .
	\end{align}
\end{subequations}
These equations can also be conveniently written in an upper-triangular matrix form~\cite{Sievert:2018imd}.    Though cumbersome,  they are straightforward to implement order by order in the opacity expansion in algebraic manipulation packages
such as Mathematica~\cite{Mathematica}.

The reaction operator Eq.~\eqref{e:reacts} encodes both the induced splitting and the momentum broadening of the total jet transverse momentum to a final observed value $\ul{p}$ from some initial value $\ul{p} - \ul{q}_{tot}$.  In the broad source approximation, one neglects the shift $\ul{q}_{tot}$ compared to the hard scale in the initial distribution $p^+ \frac{dN_0}{d^2 (p - q_{tot}) \, dp^+}$ of hard partons produced in the medium.  Then, the medium-induced splitting function can be considered to depend only on $x$ and $\ul{k}$ through the ratio:
%
$\frac{dN}{d^2 k \, dx} \equiv
	p^+ \frac{dN}{d^2 k \, dx \, d^2 p \, dp^+} /
p^+ \frac{dN_0}{d^2 p \, dp^+} $. 
%
We have developed a Mathematica code to solve the recursion relations~Eqs.~(\ref{e:reacts})  with initial conditions  Eqs.~(\ref{e:initconds}) to any order in opacity
for all of the four lowest order splitting kernels. We checked that our results recover the known expressions in the literature for both massless and massive quarks, and gluons to first order in opacity~\cite{Ovanesyan:2011kn,Kang:2016ofv}. In the soft gluon emission limit we further verified the that our results for the diagonal splitting 
functions to second order in opacity coincide with the  solutions reported in~\cite{Gyulassy:2000fs,Gyulassy:2000er}. The interested reader will find the full second order 
in opacity corrections to the medium-induced splitting kernels that we calculate here for the first time as supplementary material. 

%
\section{Results}
\label{numeric}
%

To evaluate the recursion relations Eqs. \eqref{e:reacts} for the medium-induced splitting functions, we employ the broad source approximation 
in a frame such that  the initial jet momentum relative to the direction of propagation is $p_T = 0$ and consider a particular model for the medium, assumed here to be a quark-gluon plasma with partial densities for relativistic massless quarks and gluons given by 
\begin{equation}
\rho_q = \frac{3  N_{DOF}^q}{4\pi^2} T^3(\ul{x},t) \zeta(3)\,,  \quad  \rho_g = \frac{  N_{DOF}^g}{\pi^2} T^3(\ul{x},t) \zeta(3)\, .
\end{equation} 
Here,  $T(\ul{x},t)$  is the temperature  that depends on the time $t$ and the position in the transverse plane of the collision $\ul{x}$,  $N_{DOF}^g = 16$, and for two light  quark flavors $N_f = 2$  we get  $N_{DOF}^q = 24$. With the Debye screening scale $\mu_D^2(\ul{x},t)  = g^2 T^2(\ul{x},t)  (1+N_f/6)$, the  jet on medium quasiparticle scattering cross sections~\cite{Gyulassy:1993hr} read
\begin{eqnarray} \label{e:GW}
&& \frac{d\sigma^{el}_{qq\rightarrow qq}}{d^2 q} (\ul{x},t) = \frac{1}{18 \pi^2} \frac{g^4}{\Big( q_T^2 + \mu_D^2(\ul{x},t)  \Big)^2} \,, \quad\
\frac{d\sigma^{el}_{gg\rightarrow gg}}{d^2 q} (\ul{x},t) = \frac{9}{32 \pi^2} \frac{g^4}{\Big( q_T^2 + \mu_D^2(\ul{x},t)  \Big)^2} \,, \quad \nonumber  \\
&& \frac{d\sigma^{el}_{qg\rightarrow qg}}{d^2 q} (\ul{x},t) = \frac{d\sigma^{el}_{gq\rightarrow gq}}{d^2 q} (\ul{x},t)  = 
\frac{1}{8 \pi^2} \frac{g^4}{\Big( q_T^2 + \mu_D^2(\ul{x},t)  \Big)^2} \,.
\label{e:density}
\end{eqnarray} 
We use a typical value of the in-medium coupling constant  $g = 2$ for illustration. The upper limits of the $q_T$ integrals in  Eqs.~(\ref{e:reacts}) 
are given by $q_{iT}^{\rm max} = \sqrt{\mu_{D\, i} E}$ and effective thermal mass of partons is included  
 by replacing $k_T^2 \Rightarrow k_T^2 + \frac{1}{N} \sum_{i=1}^N \mu^2_{D\, i}$.  Here, the index $i$  corresponds to position $z_i$  in the nested path length integrals along the direction of jet propagation, $N$ of them to ${\cal O}(\chi^N) $.  We also go from light-front to Cartesian coordinates, for example $z^+/E^+  = z/E$ and the $\int dz^+/ \lambda_R^+ = \int dz/ \lambda_R$.  For the upper limit of these line
integrals we take  a length much bigger than the size to the  QCD medium, such that the contributions outside of the quark-gluon plasma naturally vanish by 
$\rho \rightarrow 0$,  $\lambda_R \rightarrow \infty$.   The scattering length $\lambda_R$ we obtain as 
$1/\lambda_R = \sigma^{el}_{Rq\rightarrow Rq} \rho_q + \sigma^{el}_{Rg\rightarrow Rg} \rho_g$, where $R$ can be a quark or a gluon.

The only further input needed to describe the thermal medium and compute the in-medium splitting functions is the spacetime temperature profile along the jet propagation axis.  For an expanding quark-gluon plasma, this profile is set by using the 2+1D event-by-event viscous hydrodynamics code iEBE-VISHNU \cite{Shen:2014vra} with Monte-Carlo-Glauber initial conditions at an initial time of $\tau_0 = 0.6 \, \mathrm{fm}$.    The hydrodynamical evolution uses the S95-1 Equation of State~\cite{Huovinen:2009yb} to generate the temperature grids which are used to evaluate Eqs.~\eqref{e:density} and  the relevant medium quantities.  The jets themselves are initiated randomly according to the number density of binary collisions.

Numerical values of the splitting functions are then obtained by evaluating the integrals given in Eq.~\eqref{e:reacts} in the broad source approximation using the VEGAS algorithm~\cite{Lepage:1977sw}.  Compared to the leading order calculations, the second order calculations require integration of additional three dimensions for $z_2$ and $\ul{q}_2$. As the integration dimension becomes larger, the convergence of the VEGAS algorithm that approximates the probability distribution function of the integrand becomes slower. Therefore, in the calculation of the second order corrections we use a 10 times larger number of random samples than that used in the leading order calculation, in order to make the statistical uncertainties of the correction terms similar to those of the leading order results. In addition, the integrand of the second order correction term requires about 10 times larger number of floating point operations than the leading order calculation.

%
\begin{figure}
	\begin{center}
		\includegraphics[width= 0.7 \textwidth]{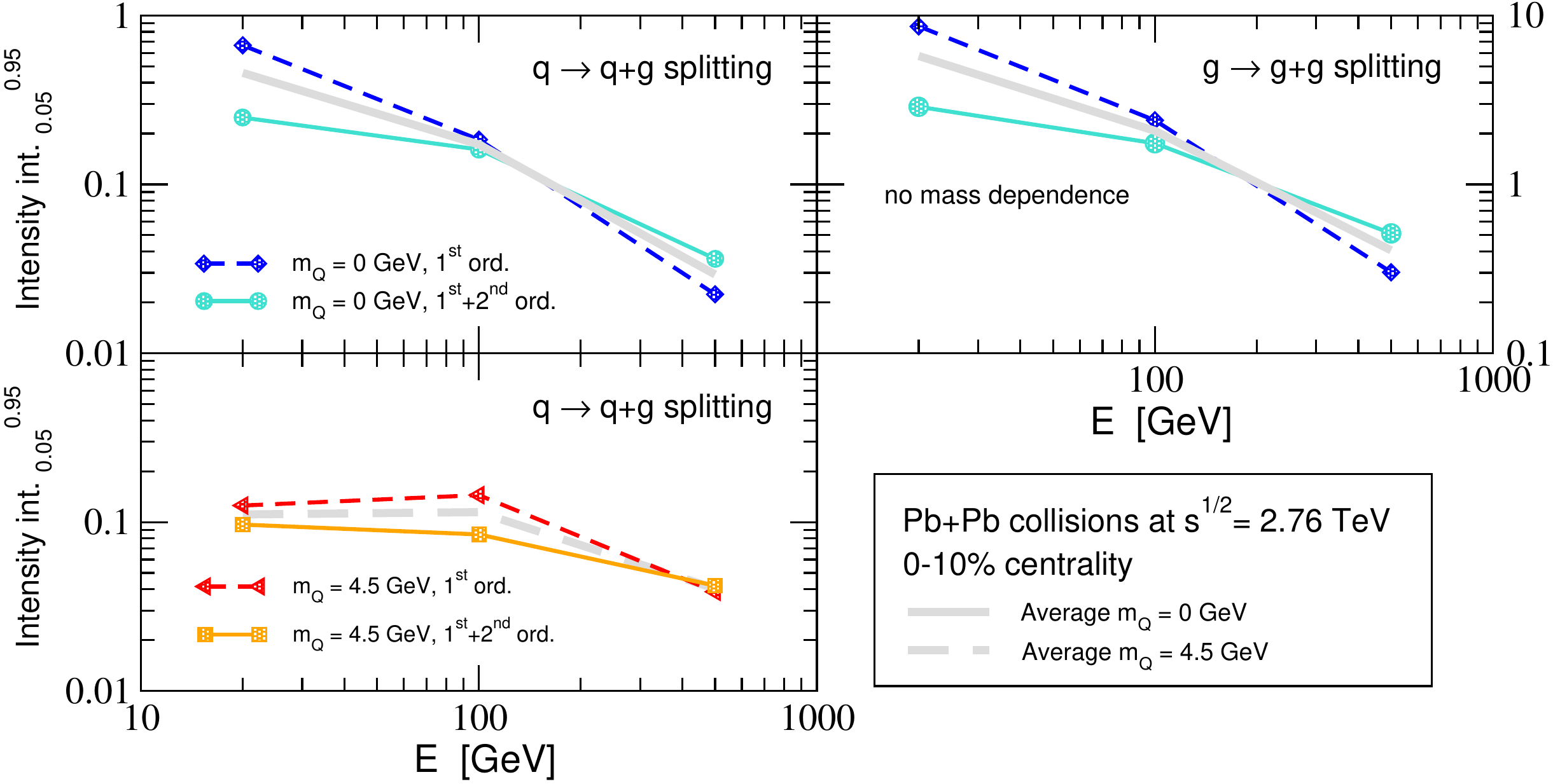} 
		\caption{The integral $\mathcal{I}^{x_\mathrm{max}}_{x_\mathrm{min}}$  Eq.~\eqref{e:intensity} for the diagonal $q\rightarrow qg$,  $g\rightarrow gg$  splitting channels as a function of jet energy in central 0-10\% Pb+Pb collisions at $\sqrt{s_{NN}} = 2.76$~TeV at the LHC.  Top panels are for light partons:  blue diamonds and turquoise circles for 1$^{\rm st}$ and 1$^{\rm st}$+2$^{\rm nd}$  orders in opacity, respectively; the bottom panel is for heavy b-quarks:  red triangles and orange squares  for 1$^{\rm st}$ and 1$^{\rm st}$+2$^{\rm nd}$  orders in opacity, respectively.  The average of the  1$^{\rm st}$ and 1$^{\rm st}$+2$^{\rm nd}$  orders in opacity for light and heavy  partons are also plotted with solid and dashed gray lines.}
		\label{f:Integrals}
	\end{center}
\end{figure}
%

We are now ready to present  our numerical results and  first  define the integral  of the splitting kernel weighed by the energy fraction of the emitted parton   over a finite range $x_\mathrm{min} \leq x \leq x_\mathrm{max}$ as follows
\begin{align} \label{e:intensity}
\mathcal{I}^{x_\mathrm{max}}_{x_\mathrm{min}} = \int_{x_\mathrm{min}}^{x_\mathrm{max}} dx \int d^2 k \:\: 
x \frac{dN}{d^2 k \, dx} ,
\end{align}
with the  transverse momentum of the splitting $ \Lambda_{QCD}^2  \leq  k_T^2  \leq 4 E^2 x(1-x) $.    
 If we restrict the energy fraction Eq.~\eqref{e:intensity} to only the diagonal channels $q \rightarrow g q$ and $g \rightarrow g g$ and the soft gluon limit $x \ll 1$, it can be interpreted as the fractional radiative energy loss $\frac{\Delta E}{E}$ of the parent parton due to the emission of soft gluons. In the general case considered above, the concept of ``parton energy loss'' is ill-defined but  $\mathcal{I}^{x_\mathrm{max}}_{x_\mathrm{min}}$ gives us a sense of the corrections due to second order in opacity to inclusive observables, such as inclusive hadron or inclusive jet suppression in  nucleus-nucleus collisions. 
Results are shown for lead-lead (Pb+Pb) collisions at $\sqrt{s_{NN}} = 2.76 \, \mathrm{TeV}$ in Fig.~\ref{f:Integrals} for the diagonal spitting channels mentioned above.  
  The integral Eq.~\eqref{e:intensity}  is plotted in Fig.~\ref{f:Integrals} for the $0-10\%$ centrality bin, using cutoffs $x_\mathrm{min} = 0.05 , x_\mathrm{max} = 0.95$ to exclude the endpoints of phase space where the energy of the partons  might be  too low to satisfy the approximation of eikonal scattering.    
  Results are shown for light quarks $m=0$,  heavy bottom quarks $m=4.5~\mathrm{GeV}$, and gluons.  We refer the reader to the figure caption for the description of the curves.  The statistical errors from the  Monte-Carlo integration are kept at  $\leq 1\%$.

%
\begin{figure}
	\begin{center}
		\includegraphics[width= 0.7\textwidth]{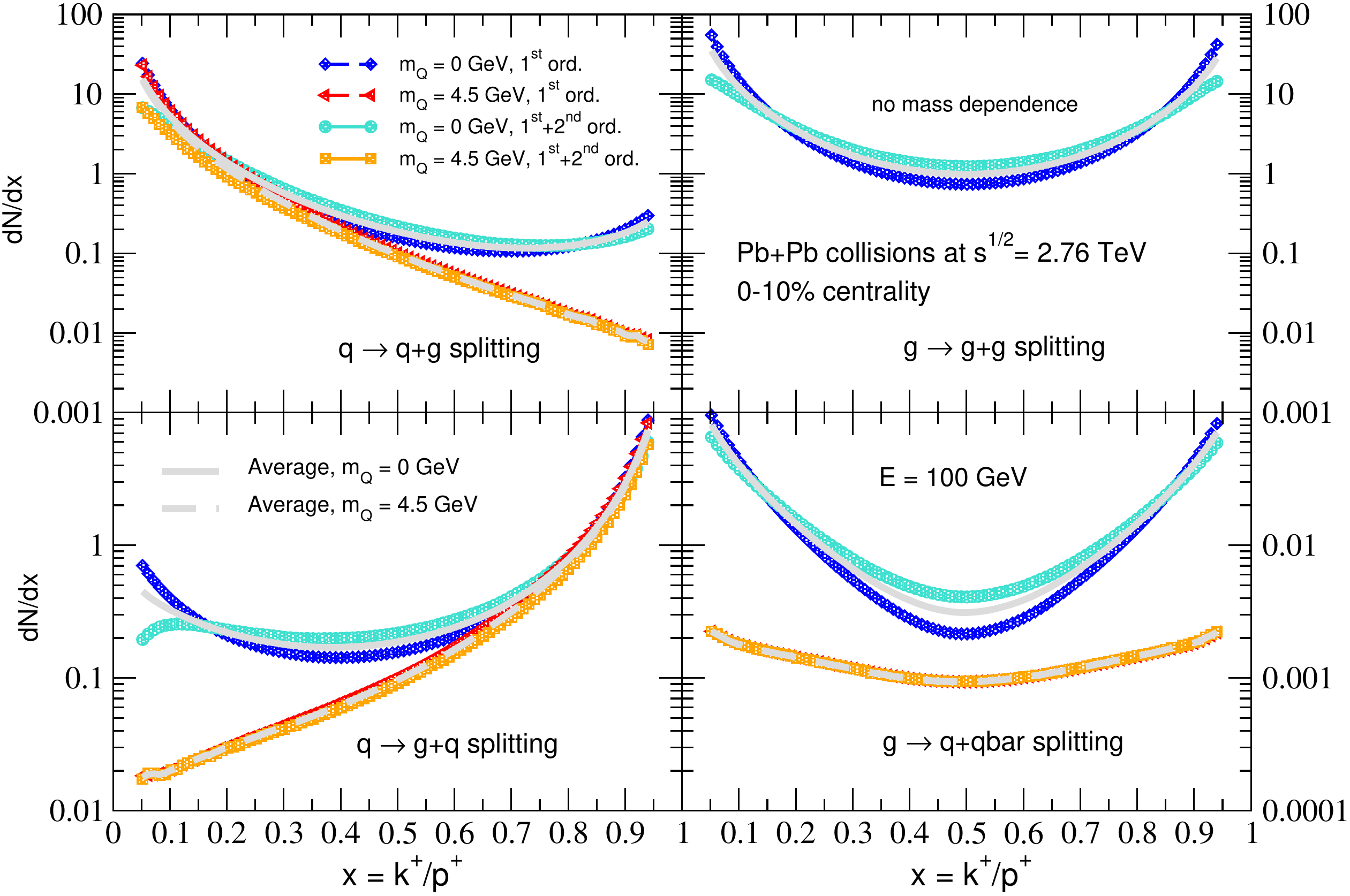} 
		\caption{The one-dimensional differential splitting functions $\frac{dN}{dx}$ for a $100 \, \mathrm{GeV}$ jet as a function of the splitting fraction $x$.   The symbol and color scheme is  the same as in Fig.~\ref{f:Integrals}, and so is the underlying QGP medium. Computations for both  the diagonal and  off-diagonal branchings are presented.}
		\label{f:Intensity}
	\end{center}
\end{figure}
%

 For light quark jets and gluon jets, a smoothing of the energy dependence is seen, with reduced in-medium shower intensity at lower jet energies and increased in-medium shower intensity at higher jet energies.  For jets with energies around $\sim 100~\mathrm{GeV}$, the corrections from second order in opacity appear to be modest, of order $\sim\mathcal{O}(10\%)$, while the corrections at both higher and lower jet energies are larger.  Based on the calculations of higher orders in opacities in the soft gluon approximation~\cite{Gyulassy:2000fs,Gyulassy:2000er,Wicks:2008ta}, the corrections tend to alternate in sign, such that the first and second orders in opacity bound the resummed result both above and below, approximating an error band associated with the opacity expansion. We show by a gray line the average of the   1$^{\rm st}$ and 1$^{\rm st}$+2$^{\rm nd}$  orders in opacity, which we expect will be suitable for 
 phenomenological implementations.  For  b-quarks the most pronounced feature remains the reduction of the in-medium shower intensity for small jet energies  due to the heavy quark mass effects, already seen in the soft gluon approximation~\cite{Dokshitzer:2001zm,Djordjevic:2003zk}.

The one-dimensional spectrum of the parton splitting  $\frac{dN}{dx}$ is shown in Fig.~\ref{f:Intensity} for a $100 \, \mathrm{GeV}$ jet.   For light partons the corrections from second order in opacity in the small $x$ and large $x$  regions are all negative and sizeable, of order $30\% - 70\%$ depending on the channel.  Interestingly, however, in all channels in the region of democratic branching $x \sim 0.5$ the second order correction is {\textit{positive}}, enhancing the branching probability and thereby producing  a smoother $x$ dependence.    We have checked that at higher jet energies this region grows considerably, covers most of the phase space in $x$ and  is entirely responsible for  enhanced in-medium shower intensity  relative to  the lowest order in opacity.  For heavy quarks 
2$^{\rm nd}$ order in opacity corrections are noticeably smaller in most parts of phase space.

%
\begin{figure}
	\begin{center}
		\includegraphics[width= 0.7\textwidth]{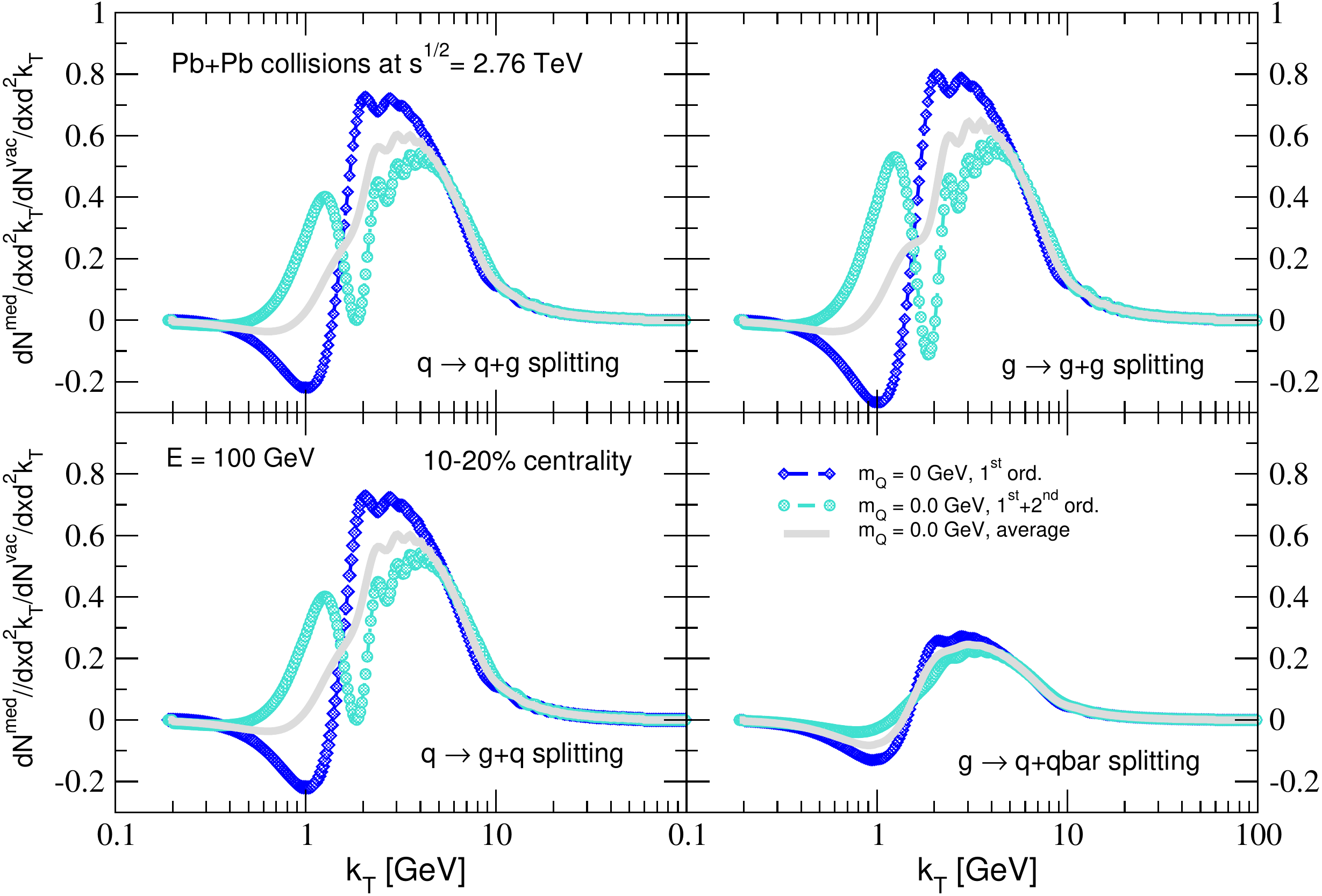} 
		\caption{The transverse momentum distribution of the medium-induced radiation, as a ratio to the vacuum radiation spectrum Eq. \eqref{e:0thOrd}.  Here the distribution is shown for a $100\, \mathrm{GeV}$ jet and $x = 0.3$.  We have chosen a $10-20\%$ centrality cut of $\sqrt{s_{NN}} = 2.76 \, \mathrm{TeV}$ PbPb collisions. The symbol and color scheme is  the same as in Fig.~\ref{f:Intensity}.}
		\label{f:KTspectra1020}
	\end{center}
\end{figure}
%

Finally, the transverse momentum distribution of the medium-induced radiation is illustrated in Fig.~\ref{f:KTspectra1020} for a $100 \, \mathrm{GeV}$ jet at $x = $0.3. We have performed simulations for several centralities and choose the $10-20\%$ centrality bin as an example.  Here, the medium-induced spectrum  $\frac{dN^{\mathrm{med}}}{d^2 k \, dx} = \sum_{n = 1}^N \left.\frac{dN}{d^2k dx}\right|_{\ord{\chi^n}}$ is expressed to $\ord{\chi^N}$  as a ratio to the vacuum radiation Eq.~\eqref{e:0thOrd} to highlight the changes induced by the medium. Note that the medium-induced part can be negative, since it excludes the vacuum contribution $\ord{\chi^0}$.  Shown in Fig.~\ref{f:KTspectra1020} are the spectra at $1^{st}$ and $1^{st} + 2^{nd}$ orders in opacity, along with the average of the two to approximate the resummed curve due to the oscillatory nature of the opacity expansion as discussed above.  At small $k_T$  the  2$^{\rm nd}$ order corrections can alter  the angular  spectrum  considerably.  At larger $k_T$ the second-order corrections are milder, reducing the peak region from first order  and gradually dying off around $k_T \geq 5 \, \mathrm{GeV}$.  All corrections for the $g \rightarrow q \bar q$ channel are much smaller and do not change the sign of the spectrum.

%
\section{Conclusions}
\label{conclusions}
%

In this Letter we presented solutions for the $q \rightarrow qg$, $g\rightarrow gg$, $q \rightarrow g q$, $g \rightarrow q\bar{q}$  in-medium splitting kernels  beyond the soft gluon approximation  
to  ${\cal O}(\alpha_s)$ and to  any desired order in opacity.  Our formalism relies on the universal features that such branching processes exhibit when they are expressed as solutions to iterative equations that build correlations between multiple scattering centers in the medium.  The differences between the splitting 
channels are conveniently captured by the light-front wavefunctions and coefficients that depend on the color representation of the partons that interact with the medium. As the first order in opacity branchings are well-documented in the literature~\cite{Ovanesyan:2011kn,Kang:2016ofv},  we only included second order in opacity corrections explicitly as supplementary material, together with the process-dependent coefficients and functions in Table~\ref{tablecoeffs}. The interested reader can recover our results and obtain even higher order in opacity corrections if desired.

We also demonstrated that the full process-dependent in-medium splitting kernels can be evaluated numerically to higher orders in opacity in a  realistic nuclear 
medium. In this Letter such a medium was exemplified by a quark-gluon plasma produced in high energy Pb+Pb collisions at the LHC.  
We  presented new numerical results which explore for the first time the corrections due to second order in opacity with exact kinematics  for all partonic branching channels in both the massive and massless quark limits.  As anticipated, we found the largest corrections for the most differential quantity, the transverse momentum spectrum of the medium-induced splitting functions, with these corrections being smoothed over as we move to the more integrated quantities.

  The derivation and numerical evaluation of higher order in opacity corrections is important and timely, as  their real application is in medium-induced splitting kernels that enter the  calculation of fixed order and  resummed observables involving reconstructed jets.  Corresponding to the changes we see in the more differential splitting functions, we anticipate that the greatest phenomenological impacts will be on the interpretation of  jet substructure measurements in heavy ion collisions~\cite{Adamczyk:2017yhe,Zharko:2017vsl,Sirunyan:2018gct,Sirunyan:2018qec,Aaboud:2019oac,Sirunyan:2018ncy,Aaboud:2018twu,Aaboud:2018hpb}.   Even in more inclusive observables, such as the suppression of jet production denoted $R_{AA}$, we anticipate the second order in opacity corrections to play a role, albeit milder. Specifically, they may be relevant for the theoretical understanding of the weak transverse momentum dependence  of the several-hundred GeV jets $R_{AA}$ observed by the ATLAS collaboration~\cite{Aaboud:2018twu}.

The splitting functions computed here are now ready for broad phenomenological applications. The very general formulation presented here is also amenable to scanning different collision systems across system size, deformation, and other parameters~\cite{Sievert:2019zjr}.  In future work, we will pursue this new program of jet phenomenology at second order in opacity to make contact with the realistic values of opacity achieved in typical heavy-ion collisions.  A future extension of this numerical analysis to the third order in opacity would also be a useful validation of the oscillatory nature of the opacity series and the use of the first and second orders as upper and lower bounds for the medium-induced splitting functions.


%
\section*{Acknowledgments}
This material is based upon work supported by the U.S. Department of Energy, Office of Science, Office of Nuclear Physics under DOE Contract  DE-AC52-06NA25396, the DOE Early Career Program and the LANL LDRD  Program.\\%

%
%



\end{document}